# A Point Cloud-Based Deep Learning Strategy for Protein-Ligand Binding Affinity Prediction

Yeji Wang[a], Shuo Wu[a], Yanwen Duan[a,b,c], and Yong Huang[a,c,*]

[a]Xiangya International Academy of Translational Medicine, Central South University, Changsha, Hunan, 410013, China

[b]Hunan Engineering Research Center of Combinatorial Biosynthesis and Natural Product Drug Discover, Changsha, Hunan, 410011, China

[c]National Engineering Research Center of Combinatorial Biosynthesis for Drug Discovery, Changsha, Hunan, 410011, China

***Corresponding author.**
E-mail address: jonghuang@csu.edu.cn (Y. Huang)

## Abstract

There is great interest to develop artificial intelligence-based protein-ligand affinity models due to their immense applications in drug discovery. In this paper, PointNet and PointTransformer, two pointwise multi-layer perceptrons have been applied for protein-ligand affinity prediction for the first time. Three-dimensional point clouds could be rapidly generated from the data sets in PDBbind-2016, which contain 3 772 and 11 327 individual point clouds derived from the refined or/and general sets, respectively. These point clouds were used to train PointNet or PointTransformer, resulting in protein-ligand affinity prediction models with Pearson correlation coefficients $R$ = 0.831 or 0.859 from the larger point clouds respectively, based on the CASF-2016 benchmark test. The analysis of the parameters suggests that the two deep learning models were capable to learn many interactions between proteins and their ligands, and these key atoms for the interaction could be visualized in point clouds. The protein-ligand interaction features learned by PointTransformer could be further adapted for the XGBoost-based machine learning algorithm, resulting in prediction models with an average $R_p$ of 0.831, which is on par with the state-of-the-art machine learning models based on PDBbind database. These results suggest that point clouds derived from the PDBbind datasets are useful to evaluate the performance of 3D point clouds-centered deep learning algorithms, which could learn critical protein-ligand interactions from natural evolution or medicinal chemistry and have wide applications in studying protein-ligand interactions.

## 1. Introduction

Space and matters in the Universe are in three-dimensional and their 3D information can be obtained, processed, stored, and utilized in a variety of ways, using depth images, meshes, and voxel grids, among others.[1] A point cloud consists of points with x, y, and z coordinates and additional Information, which is a commonly used digital format to preserve the original geometric information of the 3D objects. Many deep learning methods have been developed to process and train point clouds, which may be widely adapted in robotics, computer vision, and autonomous driving.[2] Recently, Sarupria and co-workers pioneeringly used the pointwise multi-layer perceptron PointNet to developed a generic framework for identifying local structures in molecular simulations. This method showed 99.5% accuracy for crystal structure identification in Lennard-Jones, water, and mesophase systems.[3] In two preprints, Jacobs et al. developed ASYNT-GAN, which used U-Net as an encoder and PointNet as a decoder to generate ligand structures in 3D space for *de novo* drug design,[4] while Li et al. adapted submanifold sparse convolution based U-Net to develop PointSite for the prediction of ligand-binding atoms in proteins.[5] These studies suggest the great potential of point clouds in biology and chemistry, since protein-ligand interactions may also be represented without any discretization and could be processed to point clouds for deep learning. The trained deep learning models may learn these intrinsic interaction features, governed by many non-covalent interactions, including hydrogen-bonding, π-π stacking and hydrophobic interactions, among others, which should greatly facilitate the prediction of protein-ligand affinity.

*In silico* drug design often requires accurate prediction of the affinity between a ligand (often a small molecule) and a target protein. In recent years, many mathematic models based on machine learning or deep learning have shown excellent capabilities for protein-ligand affinity prediction, and their performance exceeds traditional scoring functions (X-Score,[6] CyScore,[7] and Autodock Vina[8]) with Pearson correlation coefficient $R_p$ approaching to 0.861.[9,10] At present, there are many methods to process the 3D structural information of proteins and ligands. Their 3D structural information can be converted into molecular descriptors (FPRC, PerSpect, PSH),[11–13] expressed in 2D interaction diagrams

(DeepBindRG),[14] converted into 3D voxels ($K_{Deep}$, AK-Score, DeepAtom),[15–17] or graphic data (Graph-CNN, GraphBAR).[18,19]

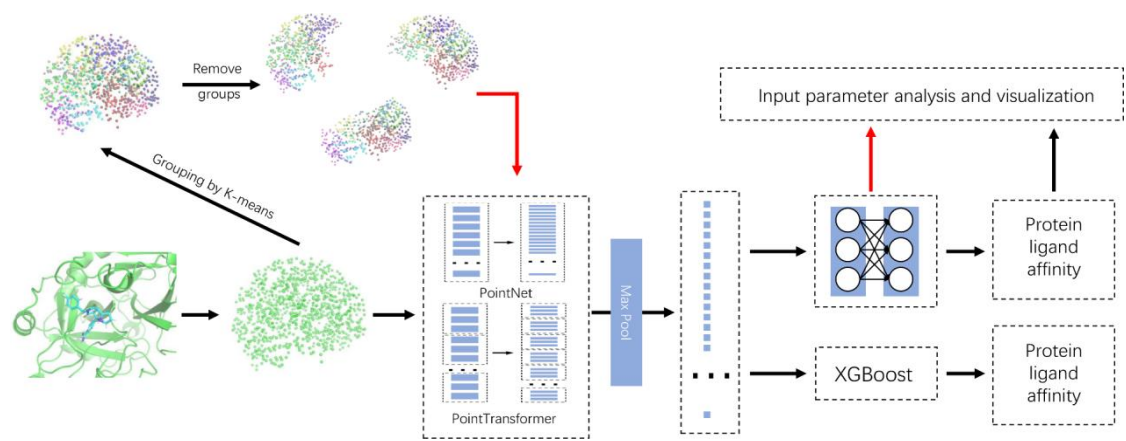

**Figure 1.** The article's research methodology.

Since the point cloud can directly input the atomic coordinates and their properties into the neural network after simple normalization, which may greatly reduce the pre-processing time of 3D protein-ligand structures. Therefore, we hypothesize that protein-ligand affinity prediction deep learning models based on point clouds may feature faster and simpler preprocessing than many of the voxelization-based approaches (**Figure 1**). Moreover, since each point in the point cloud represents an existing atom in protein-ligand structures, this salient feature would facilitate the interpretation of the developed models through visualization. In this study, we have selected two neural network structures, PointNet and PointTransformer to establish the protein-ligand affinity prediction models. After training on the PDBbind-2016 refined and a larger point clouds respectively, the models showed $R_p$ of 0.803 and 0.831 for PointNet, or 0.791 and 0.859 for PointTransfomer. These new deep learning models could learn protein-ligand interactions based on the input parameter analysis, and these key atoms for the interaction could be also visualized. Furthermore, the obtained protein-ligand interaction features from PointTransformer were adapted for the XGBoost-based machine learning algorithm, which resulted in protein-ligand affinity prediction models with $R_p$ on par with the state-of-the-art machine learning models including PerSpect ML and PSH-ML, based on benchmark tests using PDBbind-2007, PDBbind-2013, and PDBbind-2016. Our study suggests that the generated deep learning models from PDBbind-derived point clouds could learn the intrinsic interaction features between proteins and their ligands, which would have wide applications in studying protein-ligand interactions.

2. Methods

2.1 Datasets

2.1.1. Preprocessing Data

We first utilized the refined set in PDBbind-2016 as the training set in this study. The refined set contains 4 057 protein-ligand complex data sets. We employed the core set as the test set, and the remaining 3 772 data sets as the training and validation sets.[20] In addition, in order to investigate the model's performance on a larger data set, we also utilized the combination of the refined and general sets of PDBbind-2016 as a larger training set. Openbabel and Pymol are used for all the pre-processing steps.[21,22] To preprocess the PDBbind-2016 database, we adapted the following rules: (1) Entries that contain peptide ligands were deleted (590); (2) The entries containing covalently-bonded ligands were deleted (379); (3) The entries with insufficient ligands were deleted (481). All data from the test set were removed from the training set. Finally, a bigger training set of 11 327 data sets were obtained. Prior to transferring these data sets to the point cloud format, we eliminated solvents, metals, and ions from the complexes.

2.1.2. Transfer data sets from PDBbind-2016 to Point cloud formats

The atom numbers of the protein and its ligand in each data set were first counted (**Figure S1**). Considering rotation and translation should have no effect on the affinity prediction results, we mitigated their effect by aligning coordinates and rotation data.[23] Prior to incorporating the point clouds for training, we aligned the coordinates of the complex to the ligand's center, which ensured that the model was not affected by translation. We started with the 1 024 atoms closest to the ligand's center from the preprocessed data sets. To simplify computations, we considered the atoms of protein and ligand separately, while disregarded the covalent-bond relationships in proteins or ligands. Six types of atom information are contained in each point, corresponding to a single atom, including x, y, and z coordinates, Van der Waals radius, atomic weight, and their sources (1 for proteins and -1 for ligands). Atomic coordinates were normalized by the distance of the atom farthest from the ligand center. All other parameters including radius and atomic weight, were also normalized (Table S1). If the total number of atoms is fewer than 1 024, additional points with all parameters set to zero are created to compensate. In Algorithm S1, we present pseudo-code for calculating these data.

To investigate the effect of different atomic attribute inputs on the prediction results, we used two additional input methods: (1) Seven atom-type channels, including hydrogen, carbon, nitrogen, oxygen, phosphorus, sulfur, and halogen were added; (2) The sampling atoms were increased from 1 024 to 2 048.

### 2.2. Comparison of pre-processing and inference time of different models

We first generated point clouds data from the processed PDBbind data using Python and Openbabel. To further improve the performance of data processing, C++ was employed to accelerate the generation of point clouds. To compare the pre-processing speed of point clouds by Python or C++, we also employed the HTMD framework's voxelization approach, Pafnucy's pre-processing method (voxelization), and molecular descriptor methods (such as IFPScore and Persistent Spectral Hypergraph).[11,24–26] Python was used to read files, and C was used to complete the voxelization procedure in HTMD with three voxelized grids of different volumes of 15, 25, and 35 Å. Pafnucy's pre-processing program was written in Python. All methods for molecular descriptors were written in Python. In addition, we examined the inference times of PointNet, PointTransformer, and different depth convolutional neural networks. For each prediction, different sample times were used, including Pafnucy (20), ResAtom (5), PointNet (1, 5, and 24), and PointTransformer (1 and 5).[27]

All tests were run on a single thread without the use of a graphic processing unit (GPU).

### 2.3. PointNet and PointTransformer architecture and training

### 2.3.1. The neural network structures

Figure S2 depicts the PointNet structure, which comprises three modules including an encoder layer, a maximum pooling layer, and a connected layer.[23] The encoder layer consists of two transform modules and three 1-D convolution module. Each of the two transform modules contains two tiny models that are trained in conjunction with the network. The fully connected layer contains two hidden layers, while a BatchNorm1d and an activation function Rectified Linear Unit are coupled after each hidden layer.[28] There is also a dropout layer (0.5) at the end of the neural network architecture. The network structure of the PointTransformer is depicted in **Figure S3**, which also consists of three modules including an encoder layer, a maximum pooling layer, and a fully connected layer.[2] The encoder layer consists of a completely linked layer, several transformer layers, and several transition down layers. The transformer layer is comprised of two linear layers and a self-attention layer with a residual module comprise. The transition down layer consists of a completely connected sampling layer and a layer for pooling local maximal values. Given that the order of the points in the input should have no effect on the prediction outcomes, a symmetric function maximum pooling layer is added after the two models extract their features. In **Figure S4**, the principle was illustrated with a simple example.

**2.3.2. Training, Validation and Testing**

Before the above generated point clouds were trained, the point clouds from the training set and validation set were rotated 90, 180, and 270 degrees around the ligand center respectively, along the x, y, and z axes of each atom, which increases the size of the point clouds by 24 times. In the benchmark test, the final projected value was the average of 24 outputs from each point cloud. To train the point clouds of 11 327 data sets, we utilized two approaches to amplify the point clouds by firstly rotate each data set for 24 times in preparation and rotate each data set at a random angle for each epoch in training. The procedure for evaluating the pre-rotation model was identical to that for testing the refined set. The randomly rotated model outputs the mean of five random rotations. The reason for employing two different training approaches for the point clouds of 11 327 data sets was that the random rotation model can significantly minimize the number of inferences required for a single ligand. Since integrating multiple models can increase the performance of the prediction model, we integrated five best models in the final benchmark test.

All training is conducted in a Tesla V100 32G by in double precision. To train the above point clouds from PDBbind-2016, we utilized the SGD optimizer with a learning rate of 0.001 for training the data sets from the curated refined sets, while 0.003 for training the data sets from the curated refined and general sets. SmoothL1Loss was chosen as the loss function.[29]

$$loss(x,y) = \frac{1}{n}\sum_i \begin{cases} 0.5(x_i - y_i)^2, & if |x_i - y_i| < 1 \\ |x_i - y_i| - 0.5, & other \end{cases}$$

In the benchmark test, the final output result was the integrated prediction results from five best models. To evaluate the model's performance, we used the Pearson correlation coefficient ($R_P$), the Spearman correlation coefficient ($R_S$), the root mean squared error (RMSE), and the mean absolute error (MAE).

**2.4. Input parameter analysis and visualization**

To investigate whether PointNet and PointTransformer could fully explore the input factors and learn important protein-ligand interaction features rather than merely remembering the ligand binding pockets in the proteins, we eliminated individual parameter (Van der Waals radius, atomic weight, atom sources, all protein atoms, and all ligand atoms) from each data set, by replacing the removed data with 0.

We also visualized the prediction results from six data sets, including 3L7B, 4IH7, 4KZQ, 1OYT, 2YKI and 5C2H.[30-36] For each model, two best predictions, e.g., the predicted and true affinity of the protein-ligand complexes have the least difference (less 0.04), and two worst predictions, e.g., the predicted and true values of the protein-ligand complexes have the biggest differences (more than 3.70), were used for visualization. 3L7B and 3IH7 were visualized using PointNet, whereas 4KZQ and 1OYT were visualized by PointTransformer. 2YKI and 5C2H, two models performed poorly were also visualized.

We used the K-means method to cluster 1 024 points in the six point clouds data and aggregated them into 20 groups. Then, we deleted each group of data from the point cloud independently and predicted the affinity of the processed data sets using PointNet or PointTransformer. The difference between the predicted result after eliminating each group and the unremoved data was tallied. If the difference exceeds the average value, it is regarded as a critical atom affecting the prediction result. We labeled each point used with a sphere, where a distance deviation more than or equal to the average distance deviation are highlighted in red, while those with a difference less than the average deviation are highlighted in blue.

**2.5 PointTransformer features for machine learning.**

A single model trained by PointTransformer can output features with a length of 512. The PointTransformer ensemble model was used to extract features, and each point cloud data set was randomly rotated five times upon input. Consequently, a 25 × 512 interactive features were created. To minimize the effect of rotation on prediction results, we repeated the preceding procedure 30 times to produce a training set.

We employed XGBoost to construct machine learning models by using protein-ligand interaction features learned by PointTransformer. **Table 1** shows the detailed parameters for training the machine learning model. The refined sets of PDBbind-2007, PDBbind-2013, and PDBbind-2016 without the core set were used for training (**Table 2**). Finally, the three data sets would create 1 105 × 30, 2 764 × 30, and 3 772 × 30 training features, respectively. XGBoost was employed to train the three feature data sets. When performing the test set, we generated each data set for five times using the approach described above, and the average of the five prediction results was designated as the final prediction result.

**Table 1.** The setting of parameters for XGBoost machine learning models.

| Number of estimators | Subsample | Max depth | Learning rate | Colsample bytree |
|---|---|---|---|---|
| 500 | 0.8053 | 7 | 0.1 | 0.7134 |

**Table 2.** Number of data sets for different versions of PDBbind database.

| Name | Training sets | Test sets |
|---|---|---|
| PDBbind-2007 | 1 105 | 195 |
| PDBbind-2013 | 2 764 | 195 |
| PDBbind-2016 | 3 772 | 285 |

## 3. Results and Discussions

### 3.1. Comparison of pre-processing and inference time of different models

Since the order of atomic input should not affect the protein-ligand prediction results, we selected PointNet to establish protein and ligand affinity prediction model.[23] PointNet directly inputs point clouds and is immune to permutation through a symmetric function. It would learn pointwise features and extracts global features through its unique neural network structure. Since PointNet learns the hidden features in the point clouds without using complex 2D or 3D convolution, it would greatly reduce computation burden. Each point of the point clouds represents a single atom, which however only contains six types of atom properties including its 3D coordinates, and Van der Waals radius, among others. Since the bonding relationships of individual atoms were ignored in point clouds, we surmised that a very recently developed PointTransformer with local sampling capabilities and a self-attention layer could improve the model performance trained by PointNet.[2]

As illustrated in **Figure 2**, the point cloud preprocessing program significantly outperformed the others in terms of data processing speed. For example, when processing ribulose 1,5-bisphosphate carboxylase in complex with 2-carboxyarabinitol-1,5-diphosphate (1RBO) with over 40 000 atoms, its C++ version completed the task in less than 0.05 seconds, whereas other approaches run much lower, such as HTMD[37] (> 170 seconds) and the Persistent Spectral Hypergraph (PSH) (> 700 seconds).[11] For smaller proteins including the HDAC6 zinc-finger domain in complex with 3-(1,3-benzothiazol-2-yl)propanoic acid (6CEF) and the methylarginine-dependent interaction with Tudor domain in complex with methylarginine (3NTH), the point cloud preprocessing program still retains significant benefits, including a speed of 40 times that of the voxelization method (HTMD) and a speed of 20 000 times that of the PSH. All programs were run in a single thread.

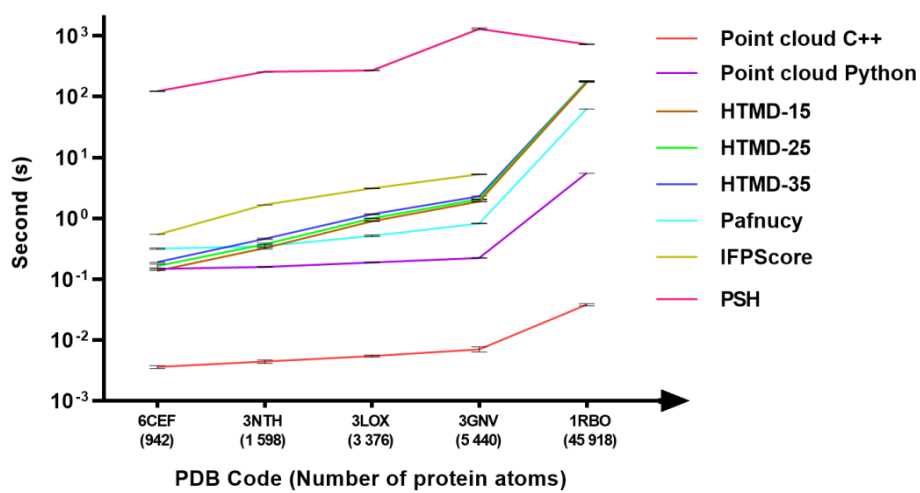

**Figure 2.** Comparison of the pre-processing time required for different scoring function. When using IFPScore for pre-processing IRBO, an error occurred. C++ and python stand for the pre-processing programing languages. PSH: Persistent Spectral Hypergraph.

In addition, we compared the amount of time required for inference using various scoring functions (**Table 3**). When PointNet used 24 samples (as tested by rotating 24 times with CASF-2016), it was still about 40% faster than Pafnucy.[25] However, we discovered that the inference time is significantly higher using PointTransformer. This could be because PointTransformer demands a great deal of time-consuming self-attention and local sampling.

**Table 3.** Comparison of the inference time among different scoring functions.

| Name | Time (s) |
| --- | --- |
| Pafnucy (20 samples) | 1.164 ± 0.03 |
| ResAtom (5 samples) | 1.043 ± 0.0306 |
| PointNet (5 samples) | 0.166 ± 0.0072 |
| PointNet (24 samples) | 0.815 ± 0.0185 |
| PointTransformer (5 samples) | 2.827 ± 0.0590 |

### 3.2. The performance of point cloud models using CASF-2016 benchmark test

We first utilized the PDBBind-2016 refined set to build several training sets with different input information of atom characteristics and sampling numbers. By examining the influence of different point cloud input approaches on PointNet, three protein-ligand affinity prediction models were generated (i.e., PointNet (1 024) and PointNet (2 048) with 1 024 and

2 048 sampling atoms respectively, PointNet (AtomChannel) with seven additional channels including hydrogen, carbon, et al). As shown in Table 4, while the MAE and RSME of PointNet (AtomChannel) were lowered from 1.43 to 1.08 or 1.74 to 1.35, no model performed better in terms of Pearson correlation coefficients. In addition, we observed that increasing sampling points from 1 024 to 2 048 slightly reduced the model performance, probably due to the increase of data noise with only limited amount of data, which would prevent it from learning the hidden features in the point cloud. Next, we chose 1 024 sampling points for the model training using PointTransformer, while the resulting model PointTransformer (1 024) ($R_p$ = 0.791) performed worse than the simpler PointNet (1 024) ($R_p$ = 0.803), as well as other the state-of-the-art models, such as $K_{deep}$, DeepAtom, PSH-GBT, and TopBP (Complex).

**Table 4.** The performance of protein-ligand binding affinity prediction on the PDBbind-2016 core set. R: the refined set of PDBbind-2016 for training; B: the use of refined and general sets of PDBbind-2016 for training. Pre-rotation is a term that refers to rotating of the point cloud for 24 times before training.

| Models | Pearson R | Spearman R | MAE | RMSE |
|---|---|---|---|---|
| PointNet-1 024 (R) | 0.803 | 0.800 | 1.43 | 1.74 |
| PointNet-AtomChannel (R) | 0.800 | 0.798 | 1.08 | 1.35 |
| PointNet-2 048 (R) | 0.785 | 0.795 | 1.48 | 1.81 |
| PointTransformer (R) | 0.791 | 0.782 | 1.09 | 1.38 |
| PointNet (B) | 0.831 | 0.827 | 0.97 | 1.26 |
| PointNet- pre-rotation (B) | 0.831 | 0.827 | 1.03 | 1.31 |
| PointTransformer (B) | 0.859 | **0.853** | **0.923** | **1.19** |
| PointTransformer-pre-rotation (B) | 0.857 | 0.850 | 0.932 | **1.19** |
| Pafnucy[25] | 0.78 | | 1.13 | 1.42 |
| $K_{deep}$[17] | 0.82 | 0.82 | | 1.27 |
| DeepAtom[16] | 0.831 | | | 1.23 |
| PSH-GBT[11] | 0.835 | | | 1.280 |
| TopBP (Complex)[9] | **0.861** | | | 1.86 |

Therefore, we trained PointNet and PointTransformer on a larger training set derived from the refined and general sets of PDBbind-2016 to generate several models (i.e., PointNet (B), PointNet-pre-rotation (B), PointTransformer (B), PointTransformer-pre-rotation (B) (Table 4). Encouragingly, the larger training set improved the performance of all generated models. For example, the performance of the PointTransformer model significantly improved over the model trained using the smaller refined set, with a $R_p$ from 0.791 to 0.859. Since deep learning would automatically learn data features through numerous iterations using basic inputs,[38,39] the model performance would likely be improved with increasing data. In machine learning algorithms, a variety of pre-designed structures are employed to extract protein-ligand interaction features and are less affected by the available data amount.[40] The PointTransformer performed better than PointNet ($R_p$ = 0.831) with a simpler network structure and the lack of local sampling capabilities, since it can learn the implicit bond information included in the atomic information by a local sampling function which is similar as a convolutional layer.[41] Additionally, PointTransformer incorporates a significant amount of self-attention, which may aid the models in learning the interaction features between proteins and their ligands.

### 3.3. Input parameter analysis

Deep learning-based protein-ligand affinity prediction methods have demonstrated excellent performances.[42] However, the majority of these studies consider these deep learning models to be *black boxes*, with only a few of them being able to explain the prediction results.[40,43] For example, Kwon et al. visualized the 3d convolutional neural network using gradient-weighted class activation mapping (Grad-CAM) and described the approximate areas that impact the major components of the affinity prediction.[15] Since each point in point clouds represents an atom from the protein-ligand complex, it would offer a unique opportunity to understand how PointNet and PointTransformer deep learning models predict the protein-ligand affinity. Table 5 summarizes the results of the input parameter analysis.

Both PointNet (B) and PointTransformer (B) relied on the entire data sets for a superior performance, since removing any data would reduce their prediction performance significantly. PointNet (B) was more sensitive to data variance than PointTransformer (B). When all the atoms belonging to proteins were deleted, PointNet's $R_p$ on CASF-2016 decreased from 0.831 to -0.2030, while PointTransformer's $R_p$ only decreased from 0.859 to 0.4495. The above input parameter analysis of the model parameters suggests that the models trained by PointNet and PointTransformer are capable of learning the interaction features between the protein and its ligand, rather than simply remembering the active pocket features of the protein and the ligand.

**Table 5.** Evaluation of the performance of PointNet (B) and PointTransforafter (B) after removing selective data in the training process.

|  | Parameter | Pearson correlation coefficient |
|---|---|---|
| PointNet (B) | All protein atom | -0.2030 |
|  | All ligand atom | 0.1803 |
|  | Atomic weight | 0.2146 |
|  | Radius | -0.1191 |
|  | Atom source (protein or ligand) | 0.1673 |
| PointTransformer(B) | All protein atom | 0.4495 |
|  | All ligand atom | 0.2014 |
|  | Atomic weight | 0.3438 |
|  | Radius | 0.3984 |
|  | Atom source (protein or ligand) | 0.1218 |

### 3.4. Point Cloud Visualization

To further investigate the protein-ligand interaction relationships learned by the trained models, we visualized four data sets with the best prediction outcomes (3L7B and 3IH7 for PointNet, 4KZQ and 1OYT for PointTransformer) and two data sets with the worst prediction outcomes (2YKI and 5C2H) (**Figures 3 and S5**). It seemed that both PointNet and PointTranfromers could detect certain atoms in at least one significant amino acid residue involved in ligand binding, but neither model recognized all the atoms in all significant amino acid residues even on the best-performing data sets. In the co-crystal structure of glycogen phosphorylase (3L7B) (**Figure 3b**), the 3′-F of 4-amino-1-(3-deoxy-3-fluoro-beta-D-glucopyranosyl) pyrimidin-2(1H)-one forms hydrogen bonds with the protein amides among A673, S674, and G675. The O-2 and O-6 of the ligand form hydrogen bonds with the protein amide close to N284 and H377, respectively, while C-5 and C-6 of the ligand have Van der Waals interactions with G135.[36]

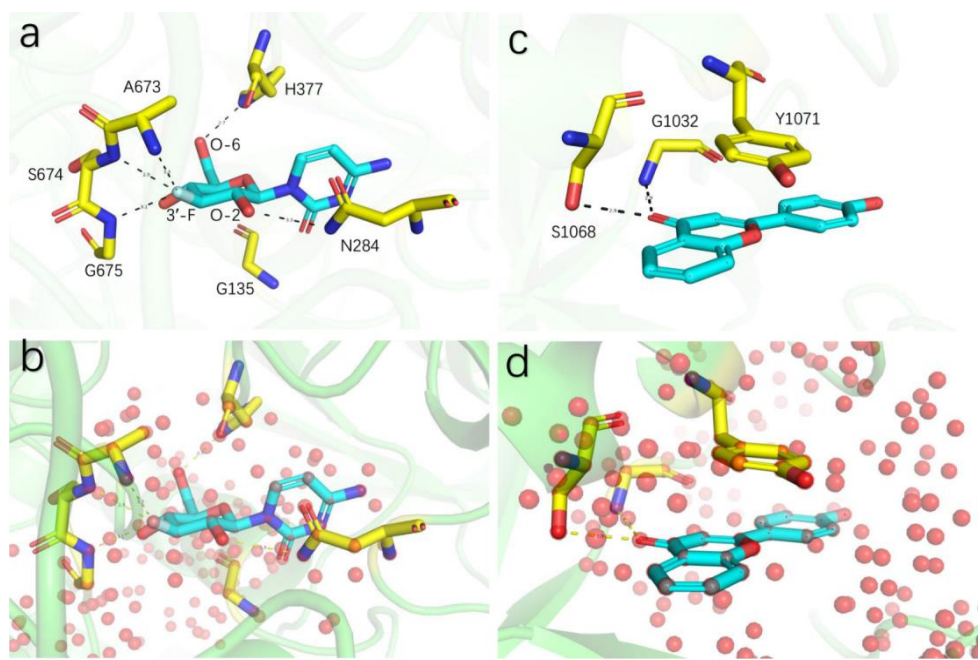

**Figure 3.** Visualization of the prediction outcomes by PointNet and PointTransformer. (a) 3L7B. (b) Visualization of 3L7B by PointNet. (c) 4KZQ. (d) Visualization of 4KZQ by PointTransformer. The blue denotes the ligand, the yellow denotes the critical amino acid residue, and the red sphere denotes the atom that has the greatest effect on the prediction outcome.

In human tankyrase (4KZQ) (**Figure 3d**), the oxygen from the benzopyran ring has π–π stacking interactions with Y1071 and hydrogen bonding interaction with the protein amide of G1032 and S1068.[35] Each importance amino acid residue contains many identified atoms. In comparison to the model trained by PointNet (**Figure S5d and S5h**), PointTransformer can detect a larger number of critical amino acids. For example, PointNet can only recognize F719 in 5C2H, while PointTransformer can recognize F719, and three additional amino acids Y683, F686, and Q716.[34]

While the two best prediction outcomes of both models in CASF-2016 benchmark are different, both models performed poorly for 2YKI and 5C2H. The elimination of the important $H_2O$ molecule to mediate ligand-protein interactions in the protein-ligand complex in 2YKI is likely one of the main reasons for the poor performance.[34] Water molecules have been shown to improve the flexibility and complementarity of proteins and ligands, as demonstrated by Wang et al. and Liu et al.[44,45] Many studies have shown that including explicit water molecules in docking simulations improves the quality of docked postures and the accuracy of binding affinity prediction.[46–50] There might also be flaws in the construction of the point cloud from 2YKI, due to the huge size of the ligand binding pocket. Some critical amino acids are too far (~ 15 Å from the ligand center) to be included in the point cloud data set (**Figure S5c and S5g**). Although PointTransformer recognized the majority of critical amino acids and PointNet only found one critical amino acid atom. (**Figure S5d and S5h**), the predicted values from the two models are significantly different from the true values. While the affinity of the ligands to the target protein in 5C2H is high, there are few data with equivalent affinity in the training sets, which therefore limited the model's effectiveness to catch the intrinsic features and is consistent with the poor performance of 5C2H in PSH machine learning.[11]

### 3.5 Features extracted PointTransformer for machine learning.

Based on the input parameter analysis and representative visualization results from point clouds, both PointNet- and PointTransformer-based models seem to learn important protein-ligand features. Therefore, we propose that these features extracted by them automatically may be used as molecular descriptors for machine learning. We therefore tested if features from PDBbind-2007, PDBbind-2013, and PDBbind-2016 retrieved by the best-performing PointTransformer model could be used. We generated 25 × 512 features from each data and repeated for 30 times. Finally, processing these data sets resulted in three training sets containing 1 105 × 30, 2 764 × 30, and 3 772 × 30 point cloud sets, respectively. The well-known XGBoost algorithm was used to construct machine learning models.[51,52] The average $R_p$ of 0.831 was obtained for the three datasets (**Table 6**). **Figure 4** illustrates a comparison of expected and experimental affinity values. We compared our model's $R_p$ to the most recent results in the literature shown in **Figure 5**.[11–13,20,25,53–57]

**Table 6.** Pearson's correlation coefficient, root mean square error and mean absolute error for XGBoost machine learning models using features extracted from the PointTransformer model on the three test sets PDBbind-2007, PDBbind-2013, and PDBbind-2016

| XGBoost models | R | RMSE | MAE |
|---|---|---|---|
| PDBbind-2007 | 0.860 | 1.27 | 1.02 |
| PDBbind-2013 | 0.802 | 1.47 | 1.17 |
| PDBbind-2016 | 0.838 | 1.22 | 0.94 |
| Average | 0.831 | 1.32 | 1.04 |

The machine learning model constructed using the PointTransformer features seemed comparable to the state-of-the-art models. Although it does not completely outperform previous machine learning models, it may have the following advantages. The procedure of obtaining features is divided into two stages: pre-processing and feature extraction using PointTransformer. Since the pre-processing time is small in comparison to the time required to extract the features, the PointTransformer extracts features regardless of the number of protein atoms. Additionally, the extracted features are more refined ($25 \times 512$) than those of PerSpect ML ($36 \times 250 \times 11 + 50 \times 100 \times 11$), allowing the description of protein-ligand interactions with fewer features. Finally, because the model was generated using Pytorch, it is simple and convenient to accelerate feature extraction using GPU-based acceleration. One limitation of PointTransformer is that its data augmentation is conducted using multiple rotations, which would increase computation time for training, in contrast to molecular descriptor-based methods capable of achieving rotation invariance.

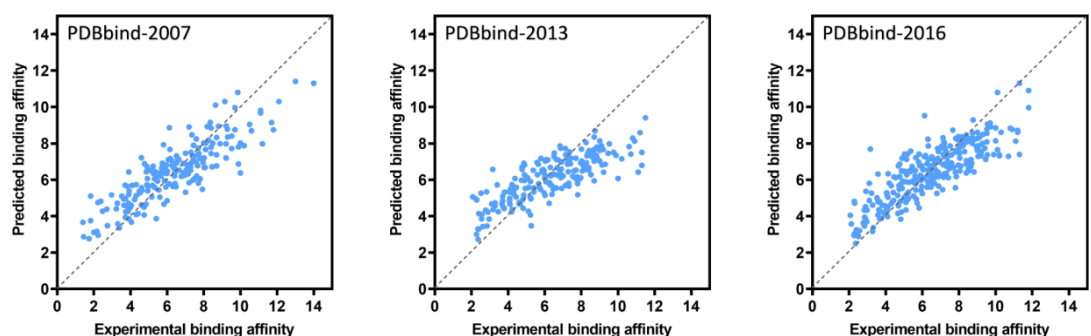

**Figure 4**. Comparison of experimental and predicted values for XGBoost models with the three test sets.

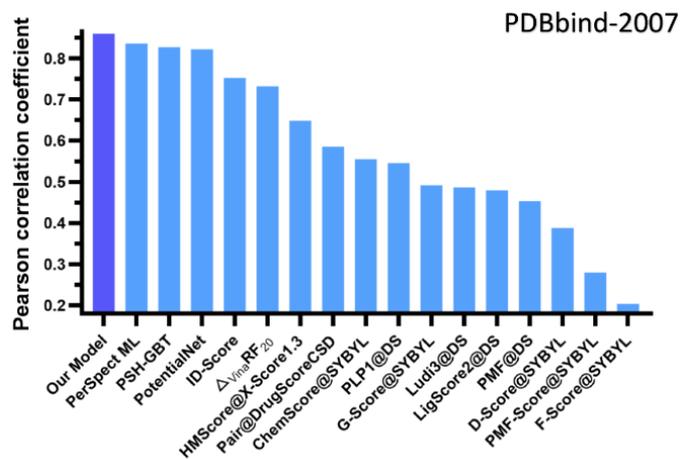

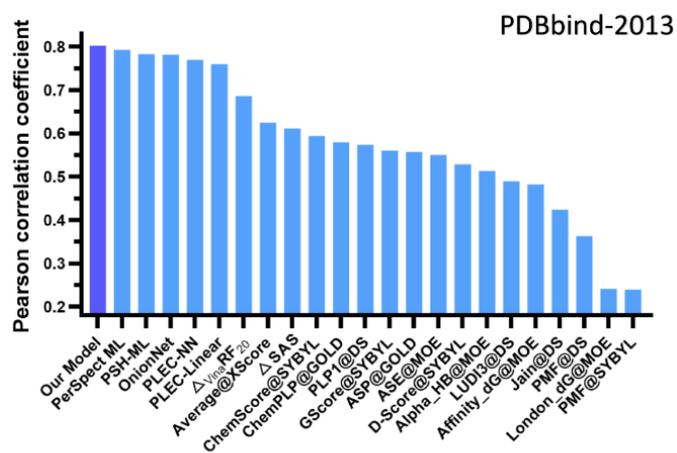

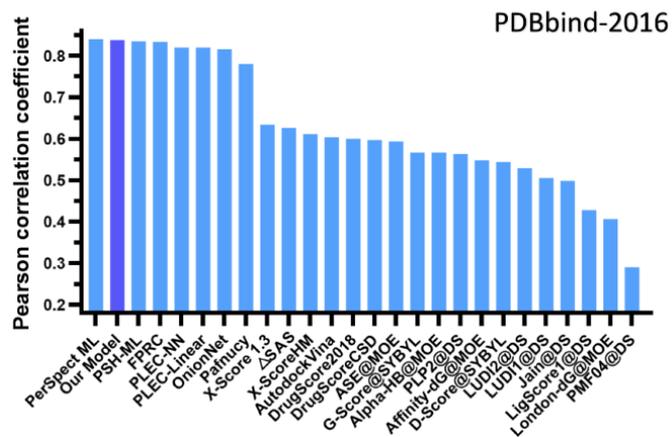

**Figure 5.** Comparison of our model with other models Pearson correlation coefficients.

## 4. Conclusion

Deep learning-based protein-ligand affinity prediction models have shown great promise in structure-based virtual screening and *in silico* drug design, while significant challenges remain including understanding the learned protein-ligand interaction features by respective models.[40] We have applied the point cloud-based neural network structures PointNet and PointTransformer for the rapid prediction of protein-ligand affinity trained on PDBbind-2016, which allowed the analysis and visualization of generated deep learning models (Table 4 and Figure 3). Further, these automatically extracted interaction descriptor from point cloud models could be adapted for the XGBoost-based machine learning algorithm, resulting in prediction models with the average of $R_p$ of 0.831 on CASF benchmark tests (Table 5). Our study suggests that point cloud-based algorithms including PointNet and PointTransformer may generate transparent and interpretable protein-ligand affinity deep learning models, which should greatly facilitate the further development of prediction models for protein-ligand interactions.

When compared to machine learning or other deep learning models, the point cloud-based models outperform them by several orders of magnitude in terms of pre-processing speed for the tested data sets (Figure 1). After training of PointNet and PointTransformer with the PDBbind data sets, the generated deep learning models have already possessed some feature extraction capabilities, which may be further trained on fresh datasets as migration models and hence lower the amount of data required.[58–60] Since the interaction features between a protein and its ligand may be effectively retrieved using PointNet and PointTransformer, these models could be used to predict critical amino acid residues important for the binding affinity of the ligand.

While the PointTransformer model achieved the superior performance, its computational cost is 15 times that of PointNet. Therefore, the performance of PointNet model could be enhanced by including more bonding information of neighboring atoms and hyperparameter optimizing, since we have only included limited information on the atom itself (i.e., atomic radius and mass). The highly efficient PointNet may be used as an encoder module to describe the state of protein-ligand complex, allowing point clouds to be applied to domains such as reinforcement learning.[61,62] These point cloud models can also be developed into a variety of other applications, including enzyme active site search and molecular docking.[39,42,63,64]

**Key Points:**

- Deep learning-based point clouds using PointNet and PointTransformer were firstly applied for the prediction of protein-ligand affinity.
- Visualization of protein-ligand interactions learned by point cloud models on the atomic level.
- Develop an XGBoost machine learning model using the features automatically retrieved from the PointTransformer models for the prediction of ligand binding affinity, comparable to the state-of-the-art scoring functions using CASF benchmark tests (2007, 2013, and 2016).

**Data and code availability:**

Data and code can be found from this link https://github.com/wyji001/Point-Cloud.

**Acknowledgment**

This work was supported in part by the High Performance Computing Center of Central South University.

**Funding**

This work was supported by the NSFC Grant 81473124 (to Y. H.); the Chinese Ministry of Education 111 Project BP0820034 (to Y. D.) and the Fundamental Research Funds for the Central Universities of Central South University (CSU) (to Y. W.).

**References**

(1) Guo, Y.; Wang, H.; Hu, Q.; Liu, H.; Liu, L.; Bennamoun, M. Deep Learning for 3d Point Clouds: A Survey. *IEEE Trans. Pattern Anal. Mach. Intell.* **2020**.

(2) Zhao, H.; Jiang, L.; Jia, J.; Torr, P.; Koltun, V. Point Transformer. *arXiv Prepr. arXiv2012.09164* **2020**.

(3) DeFever, R. S.; Targonski, C.; Hall, S. W.; Smith, M. C.; Sarupria, S. A Generalized Deep Learning Approach for Local Structure Identification in Molecular Simulations. *Chem. Sci.* **2019**, *10* (32), 7503–7515.

(4) Jacobs, I.; Maragoudakis, M. De Novo Drug Design Using Artificial Intelligence ASYNT-GAN. **2020**.

(5) Li, Z.; Yan, X.; Wei, Q.; Gao, X.; Wang, S.; Cui, S. PointSite: A Point Cloud Segmentation Tool for Identification of Protein Ligand Binding Atoms. *bioRxiv* **2019**, 831131.

(6) Wang, R.; Lai, L.; Wang, S. Further Development and Validation of Empirical Scoring Functions for Structure-Based Binding Affinity Prediction. *J. Comput. Aided. Mol. Des.* **2002**, *16* (1), 11–26.

(7) Cao, Y.; Li, L. Improved Protein–Ligand Binding Affinity Prediction by Using a Curvature-Dependent Surface-Area Model. *Bioinformatics* **2014**, *30* (12), 1674–1680.

(8) Trott, O.; Olson, A. J. AutoDock Vina: Improving the Speed and Accuracy of Docking with a New Scoring Function, Efficient Optimization, and Multithreading. *J. Comput. Chem.* **2010**, *31* (2), 455–461.

(9) Cang, Z.; Mu, L.; Wei, G.-W. Representability of Algebraic Topology for Biomolecules in Machine Learning Based Scoring and Virtual Screening. *PLoS Comput. Biol.* **2018**, *14* (1), e1005929.

(10) Li, H.; Sze, K.; Lu, G.; Ballester, P. J. Machine-learning Scoring Functions for Structure-based Virtual Screening. *Wiley Interdiscip. Rev. Comput. Mol. Sci.* **2021**, *11* (1), e1478.

(11) Liu, X.; Feng, H.; Wu, J.; Xia, K. Persistent Spectral Hypergraph Based Machine Learning (PSH-ML) for Protein-Ligand Binding Affinity Prediction. *Brief. Bioinform.* **2021**.

(12) Meng, Z.; Xia, K. Persistent Spectral–Based Machine Learning (PerSpect ML) for Protein-Ligand Binding Affinity Prediction. *Sci. Adv.* **2021**, *7* (19), eabc5329.

(13) Wee, J.; Xia, K. Forman Persistent Ricci Curvature (FPRC)-Based Machine Learning Models for Protein–Ligand Binding Affinity Prediction. *Brief. Bioinform.* **2021**.

(14) Zhang, H.; Liao, L.; Saravanan, K. M.; Yin, P.; Wei, Y. DeepBindRG: A Deep Learning Based Method for Estimating Effective Protein–Ligand Affinity. *PeerJ* **2019**, *7*, e7362.

(15) Kwon, Y.; Shin, W.-H.; Ko, J.; Lee, J. AK-Score: Accurate Protein-Ligand Binding Affinity Prediction Using an Ensemble of 3D-Convolutional Neural Networks. *Int. J. Mol. Sci.* **2020**, *21* (22), 8424.

(16) Rezaei, M. A.; Li, Y.; Wu, D. O.; Li, X.; Li, C. Deep Learning in Drug Design: Protein-Ligand Binding Affinity Prediction. *IEEE/ACM Trans. Comput. Biol. Bioinforma.* **2020**. https://doi.org/10.1109/TCBB.2020.3046945.

(17) Jiménez, J.; Skalic, M.; Martinez-Rosell, G.; De Fabritiis, G. K Deep: Protein–Ligand Absolute Binding Affinity Prediction via 3d-Convolutional Neural Networks. *J. Chem. Inf. Model.* **2018**, *58* (2), 287–296.

(18) Torng, W.; Altman, R. B. Graph Convolutional Neural Networks for Predicting Drug-Target Interactions. *J. Chem. Inf. Model.* **2019**, *59* (10), 4131–4149.

(19) Son, J.; Kim, D. Development of a Graph Convolutional Neural Network Model for Efficient Prediction of Protein-Ligand Binding Affinities. *PLoS One* **2021**, *16* (4), e0249404.

(20) Su, M.; Yang, Q.; Du, Y.; Feng, G.; Liu, Z.; Li, Y.; Wang, R. Comparative Assessment of Scoring Functions: The CASF-2016 Update. *J. Chem. Inf. Model.* **2018**, *59* (2), 895–913.

(21) O'Boyle, N. M.; Banck, M.; James, C. A.; Morley, C.; Vandermeersch, T.; Hutchison, G. R. Open Babel: An Open Chemical Toolbox. *J. Cheminform.* **2011**, *3* (1), 1–14.

(22) DeLano, W. L. The PyMOL Molecular Graphics System. *http//www. pymol. org* **2002**.

(23) Qi, C. R.; Su, H.; Mo, K.; Guibas, L. J. Pointnet: Deep Learning on Point Sets for 3d Classification and Segmentation. In *Proceedings of the IEEE conference on computer vision and pattern recognition*; 2017; pp 652–660.

(24) Wang, D. D.; Xie, H.; Yan, H. Proteo-Chemometrics Interaction Fingerprints of Protein–Ligand Complexes Predict Binding Affinity. *Bioinformatics* **2021**.

(25) Stepniewska-Dziubinska, M. M.; Zielenkiewicz, P.; Siedlecki, P. Development and Evaluation of a Deep Learning Model for Protein–Ligand Binding Affinity Prediction. *Bioinformatics* **2018**, *34* (21), 3666–3674.

(26) Doerr, S.; Harvey, M. J.; Noé, F.; De Fabritiis, G. HTMD: High-Throughput Molecular Dynamics for Molecular Discovery. *J. Chem. Theory Comput.* **2016**, *12* (4), 1845–1852.

(27) Wang, Y.; Wu, S.; Duan, Y.; Huang, Y. ResAtom System: Protein and Ligand Affinity Prediction Model Based on Deep Learning. *arXiv Prepr. arXiv2105.05125* **2021**.

(28) Nair, V.; Hinton, G. E. Rectified Linear Units Improve Restricted Boltzmann Machines. In *Icml*; 2010.

(29) Girshick, R. Fast R-Cnn. In *Proceedings of the IEEE international conference on computer vision*; 2015; pp 1440–1448.

(30) Olsen, J. A.; Banner, D. W.; Seiler, P.; Obst Sander, U.; D'Arcy, A.; Stihle, M.; Müller, K.; Diederich, F. A Fluorine Scan of Thrombin Inhibitors to Map the Fluorophilicity/Fluorophobicity of an Enzyme Active Site: Evidence for C   F··· C   O Interactions. *Angew. Chemie Int. Ed.* **2003**, *42* (22), 2507–2511.

(31) Makthal, N.; Rastegari, S.; Sanson, M.; Ma, Z.; Olsen, R. J.; Helmann, J. D.; Musser, J. M.; Kumaraswami, M. Crystal Structure of Peroxide Stress Regulator from Streptococcus Pyogenes Provides Functional Insights into the Mechanism of Oxidative Stress Sensing. *J. Biol. Chem.* **2013**, *288* (25), 18311–18324.

(32) Talamas, F. X.; Ao-Ieong, G.; Brameld, K. A.; Chin, E.; de Vicente, J.; Dunn, J. P.; Ghate, M.; Giannetti, A. M.; Harris, S. F.; Labadie, S. S. De Novo Fragment Design: A Medicinal Chemistry Approach to Fragment-Based Lead Generation. *J. Med. Chem.* **2013**, *56* (7), 3115–3119.

(33) Vallée, F.; Carrez, C.; Pilorge, F.; Dupuy, A.; Parent, A.; Bertin, L.; Thompson, F.; Ferrari, P.; Fassy, F.; Lamberton, A. Tricyclic Series of Heat Shock Protein 90 (Hsp90) Inhibitors Part I: Discovery of Tricyclic Imidazo [4, 5-c] Pyridines as Potent Inhibitors of the Hsp90 Molecular Chaperone. *J. Med. Chem.* **2011**, *54* (20), 7206–7219.

(34) Shipe, W. D.; Sharik, S. S.; Barrow, J. C.; McGaughey, G. B.; Theberge, C. R.; Uslaner, J. M.; Yan, Y.; Renger, J. J.; Smith, S. M.; Coleman, P. J. Discovery and Optimization of a Series of Pyrimidine-Based Phosphodiesterase 10A (PDE10A) Inhibitors through Fragment Screening, Structure-Based Design, and Parallel Synthesis. *J. Med. Chem.* **2015**, *58* (19), 7888–7894.

(35) Narwal, M.; Koivunen, J.; Haikarainen, T.; Obaji, E.; Legala, O. E.; Venkannagari, H.; Joensuu, P.; Pihlajaniemi, T.; Lehtio, L. Discovery of Tankyrase Inhibiting Flavones with Increased Potency and Isoenzyme Selectivity. *J. Med. Chem.* **2013**, *56* (20), 7880–7889.

(36) Tsirkone, V. G.; Tsoukala, E.; Lamprakis, C.; Manta, S.; Hayes, J. M.; Skamnaki, V. T.; Drakou, C.; Zographos, S. E.; Komiotis, D.; Leonidas, D. D. 1-(3-Deoxy-3-Fluoro-β-D-Glucopyranosyl) Pyrimidine Derivatives as Inhibitors of Glycogen Phosphorylase b: Kinetic, Crystallographic and Modelling Studies. *Bioorg. Med. Chem.* **2010**, *18* (10), 3413–3425.

(37) Jiménez, J.; Doerr, S.; Martínez-Rosell, G.; Rose, A. S.; De Fabritiis, G. DeepSite: Protein-Binding Site Predictor Using 3D-Convolutional Neural Networks. *Bioinformatics* **2017**, *33* (19), 3036–3042.

(38) Yang, J.; Shen, C.; Huang, N. Predicting or Pretending: Artificial Intelligence for Protein-Ligand Interactions Lack of Sufficiently Large and Unbiased Datasets. *Front. Pharmacol.* **2020**, *11*, 69.

(39) Heo, L.; Janson, G.; Feig, M. Physics-Based Protein Structure Refinement in the Era of Artificial Intelligence. *Proteins Struct. Funct. Bioinforma.*

(40) Qin, T.; Zhu, Z.; Wang, X. S.; Xia, J.; Wu, S. Computational Representations of Protein-Ligand Interfaces for Structure-Based Virtual Screening. *Expert Opin. Drug Discov.* **2021**, No. just-accepted.

(41) Qi, C. R.; Yi, L.; Su, H.; Guibas, L. J. Pointnet++: Deep Hierarchical Feature Learning on Point Sets in a Metric Space. *arXiv Prepr. arXiv1706.02413* **2017**.

(42) Rifaioglu, A. S.; Cetin Atalay, R.; Cansen Kahraman, D.; Doğan, T.; Martin, M.; Atalay, V. MDeePred: Novel Multi-Channel Protein Featurization for Deep Learning-Based Binding Affinity Prediction in Drug Discovery. *Bioinformatics* **2021**, *37* (5), 693–704.

(43) Zhang, Q.; Zhu, S.-C. Visual Interpretability for Deep Learning: A Survey. *arXiv Prepr. arXiv1802.00614* **2018**.

(44) Lu, Y.; Wang, R.; Yang, C.-Y.; Wang, S. Analysis of Ligand-Bound Water Molecules in High-Resolution Crystal Structures of Protein− Ligand Complexes. *J. Chem. Inf. Model.* **2007**, *47* (2), 668–675.

(45) Wong, S. E.; Lightstone, F. C. Accounting for Water Molecules in Drug Design. *Expert Opin. Drug Discov.* **2011**, *6* (1), 65–74.

(46) Dou, X.; Jiang, L.; Wang, Y.; Jin, H.; Liu, Z.; Zhang, L. Discovery of New GSK-3β Inhibitors through Structure-Based Virtual Screening. *Bioorg. Med. Chem. Lett.* **2018**, *28* (2), 160–166.

(47) Zhong, H.; Wang, Z.; Wang, X.; Liu, H.; Li, D.; Liu, H.; Yao, X.; Hou, T. Importance of a Crystalline Water Network in Docking-Based Virtual Screening: A Case Study of BRD4. *Phys. Chem. Chem. Phys.* **2019**, *21* (45), 25276–25289.

(48) Thilagavathi, R.; Mancera, R. L. Ligand− Protein Cross-Docking with Water Molecules. *J. Chem. Inf. Model.* **2010**, *50* (3), 415–421.

(49) García-Sosa, A. T.; Mancera, R. L.; Dean, P. M. WaterScore: A Novel Method for Distinguishing between Bound and Displaceable Water Molecules in the Crystal Structure of the Binding Site of Protein-Ligand Complexes. *J. Mol. Model.* **2003**, *9* (3), 172–182.

(50) Roberts, B. C.; Mancera, R. L. Ligand− Protein Docking with Water Molecules. *J. Chem. Inf. Model.* **2008**, *48* (2), 397–408.

(51) Friedman, J.; Hastie, T.; Tibshirani, R. Additive Logistic Regression: A Statistical View of Boosting (with Discussion and a Rejoinder by the Authors). *Ann. Stat.* **2000**, *28* (2), 337–407.

(52) Friedman, J. H. Greedy Function Approximation: A Gradient Boosting Machine. *Ann. Stat.* **2001**, 1189–1232.

(53) Wang, C.; Zhang, Y. Improving Scoring-docking-screening Powers of Protein–Ligand Scoring Functions Using Random Forest. *J. Comput. Chem.* **2017**, *38* (3), 169–177.

(54) Zheng, L.; Fan, J.; Mu, Y. Onionnet: A Multiple-Layer Intermolecular-Contact-Based Convolutional Neural Network for Protein–Ligand Binding Affinity Prediction. *ACS omega* **2019**, *4* (14), 15956–15965.

(55) Li, G.-B.; Yang, L.-L.; Wang, W.-J.; Li, L.-L.; Yang, S.-Y. ID-Score: A New Empirical Scoring Function Based on a Comprehensive Set of Descriptors Related to Protein–Ligand Interactions. *J. Chem. Inf. Model.* **2013**, *53* (3), 592–600.

(56) Wójcikowski, M.; Kukiełka, M.; Stepniewska-Dziubinska, M. M.; Siedlecki, P. Development of a Protein–Ligand Extended Connectivity (PLEC) Fingerprint and Its Application for Binding Affinity Predictions. *Bioinformatics* **2019**, *35* (8), 1334–1341.

(57) Feinberg, E. N.; Sur, D.; Wu, Z.; Husic, B. E.; Mai, H.; Li, Y.; Sun, S.; Yang, J.; Ramsundar, B.; Pande, V. S. PotentialNet for Molecular Property Prediction. *ACS Cent. Sci.* **2018**, *4* (11), 1520–1530.

(58) Ghasemi, F.; Fassihi, A.; Pérez-Sánchez, H.; Mehri Dehnavi, A. The Role of Different Sampling Methods in Improving Biological Activity Prediction Using Deep Belief Network. *J. Comput. Chem.* **2017**, *38* (4), 195–203.

(59) Chen, H.; Engkvist, O.; Wang, Y.; Olivecrona, M.; Blaschke, T. The Rise of Deep Learning in Drug Discovery. *Drug Discov. Today* **2018**, *23* (6), 1241–1250.

(60) Cai, C.; Wang, S.; Xu, Y.; Zhang, W.; Tang, K.; Ouyang, Q.; Lai, L.; Pei, J. Transfer Learning for Drug Discovery. *J. Med. Chem.* **2020**, *63* (16), 8683–8694.

(61) Serrano, A.; Imbernon, B.; Perez-Sanchez, H.; Cecilia, J. M.; Bueno-Crespo, A.; Abellan, J. L. QN-Docking: An Innovative Molecular Docking Methodology Based on Q-Networks. *Appl. Soft Comput.* **2020**, *96*, 106678.

(62) Jose, J.; Gupta, K.; Alam, U.; Jatana, N.; Arora, P. Reinforcement Learning Based Approach for Ligand Pose Prediction. *bioRxiv* **2021**.

(63) Mylonas, S. K.; Axenopoulos, A.; Daras, P. DeepSurf: A Surface-Based Deep Learning Approach for the Prediction of Ligand Binding Sites on Proteins. *arXiv Prepr. arXiv2002.05643* **2020**.

(64) McNutt, A. T.; Francoeur, P.; Aggarwal, R.; Masuda, T.; Meli, R.; Ragoza, M.; Sunseri, J.; Koes, D. R. GNINA 1.0: Molecular Docking with Deep Learning. *J. Cheminform.* **2021**, *13* (1). https://doi.org/10.1186/s13321-021-00522-2.

# Supporting Information

## A Point Cloud-Based Deep Learning Strategy for Protein-Ligand Binding Affinity Prediction

Yeji Wang[1], Shuo Wu[1,2], Yanwen Duan[1,2,3] and Yong Huang[1,3,*]

[1] Xiangya International Academy of Translational Medicine, Central South University, Changsha, Hunan, 410013, China.

[2] Hunan Engineering Research Center of Combinatorial Biosynthesis and Natural Product Drug Discover, Changsha, Hunan, 410011, China.

[3] National Engineering Research Center of Combinatorial Biosynthesis for Drug Discovery, Changsha, Hunan, 410011, China

*To whom correspondence should be addressed.

**Table S1.** Normalized data for different atoms

| Atomic number | atomic weight | radius |
| --- | --- | --- |
| 1 | -0.929 | 0.55 |
| 6 | -0.642 | 0.85 |
| 7 | -0.590 | 0.775 |
| 8 | -0.538 | 0.76 |
| 9 | -0.460 | 0.735 |
| 15 | -0.172 | 0.90 |
| 16 | -0.120 | 0.90 |
| 17 | -0.030 | 0.875 |
| 35 | 1.129 | 0.915 |
| 53 | 2.356 | 0.99 |
| Other | 0 | 0 |

**Table S2.** Hyperparameter Optimization for PointNet

| Optimizer | Batch Size | Learning rate | R |
|---|---|---|---|
| Adam | 128 | 0.01 | 0.725 |
|  |  | 0.001 | 0.728 |
|  |  | 0.0001 | 0.715 |
| SGD |  | 0.1 | 0.715 |
|  |  | 0.01 | 0.742 |
|  |  | 0.001 | 0.700 |
|  |  | 0.0001 | 0.729 |
|  | 64 | 0.001 | 0.757 |
|  | 32 | 0.001 | 0.753 |

**Table S3.** The performance of single model.

| Name | 1 | 2 | 3 | 4 | 5 | Mean |
|---|---|---|---|---|---|---|
| PointScore-1024(R) | 0.762 | 0.759 | 0.754 | 0.753 | 0.751 | 0.756 |
| PointScore-AtomChannel(R) | 0.772 | 0.770 | 0.769 | 0.767 | 0.748 | 0.765 |
| PointScore-2048(R) | 0.755 | 0.751 | 0.748 | 0.747 | 0.746 | 0.749 |
| PointTransformer(R) | 0.760 | 0.758 | 0.754 | 0.750 | 0.746 | 0.753 |
| PointScore(B) | 0.794 | 0.791 | 0.788 | 0.787 | 0.764 | 0.784 |
| PointScore-prerot(B) | 0.798 | 0.790 | 0.789 | 0.788 | 0.768 | 0.787 |
| PointTransformer(B) | 0.827 | 0.822 | 0.821 | 0.820 | 0.819 | 0.822 |
| PointTransformer-prerot(B) | 0.845 | 0.835 | 0.831 | 0.826 | 0.825 | 0.832 |

**Algorithm S1.** pseudo-code computation

**Input:** Protein and Ligand structure

1: Align the coordinates with the center of the ligand

2: Calculate the distance of all atoms from the center

3: Get all the atoms of ligand and (1024- ligand atom number) protein atoms closest to the center.

4: **for** each selected atom **do**

5:     a, b, c ← AtomCoords, AtomRadius, AtomWeight

6:     **if** atom **from** protein

7:         d ← 1

8:     **elif** atom **from** ligand

9:         d ← -1

10:    **end if**

11:    Point Clouds [ i ] ← [a, b, c, d]

12: **end for**

**Output:** Point Clouds

**Figure S1**. Distribution of atomic number of big data and casf-2016

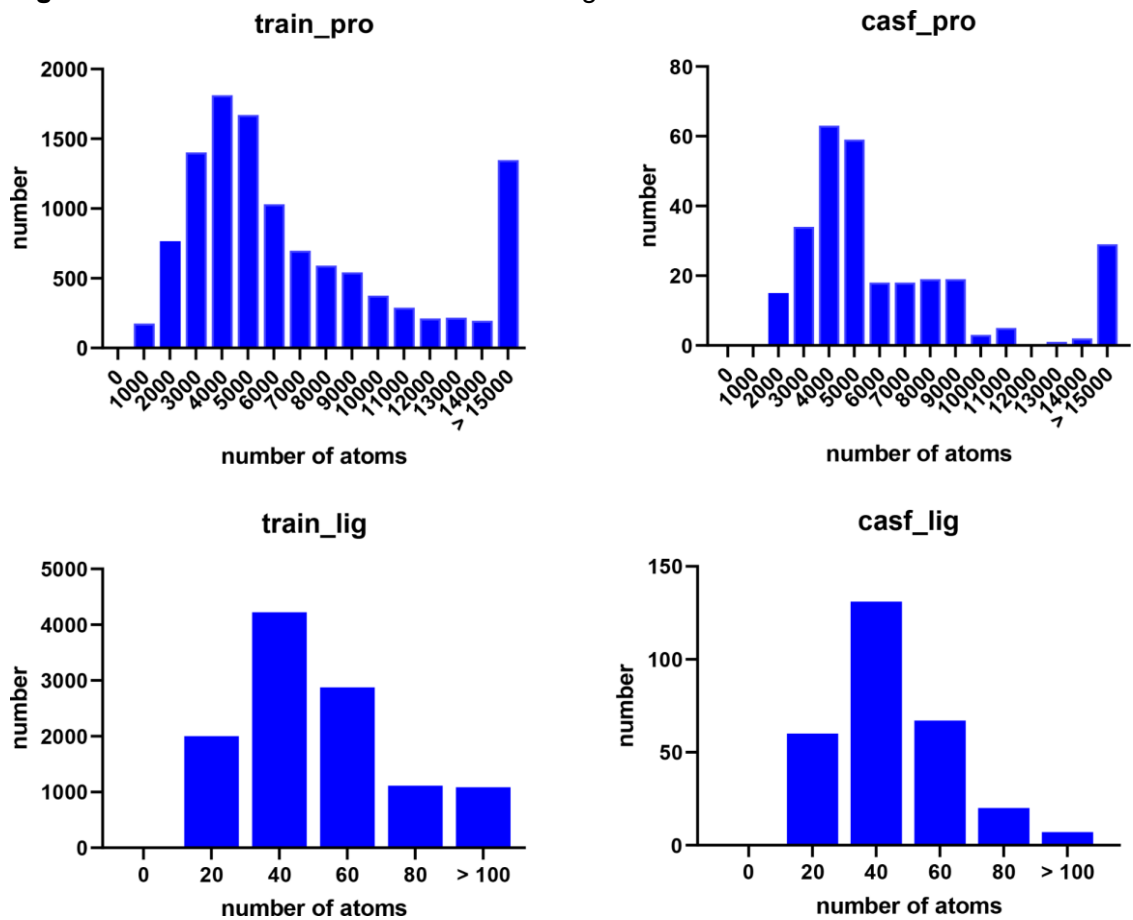

**Figure S2. PointNet architecture.**

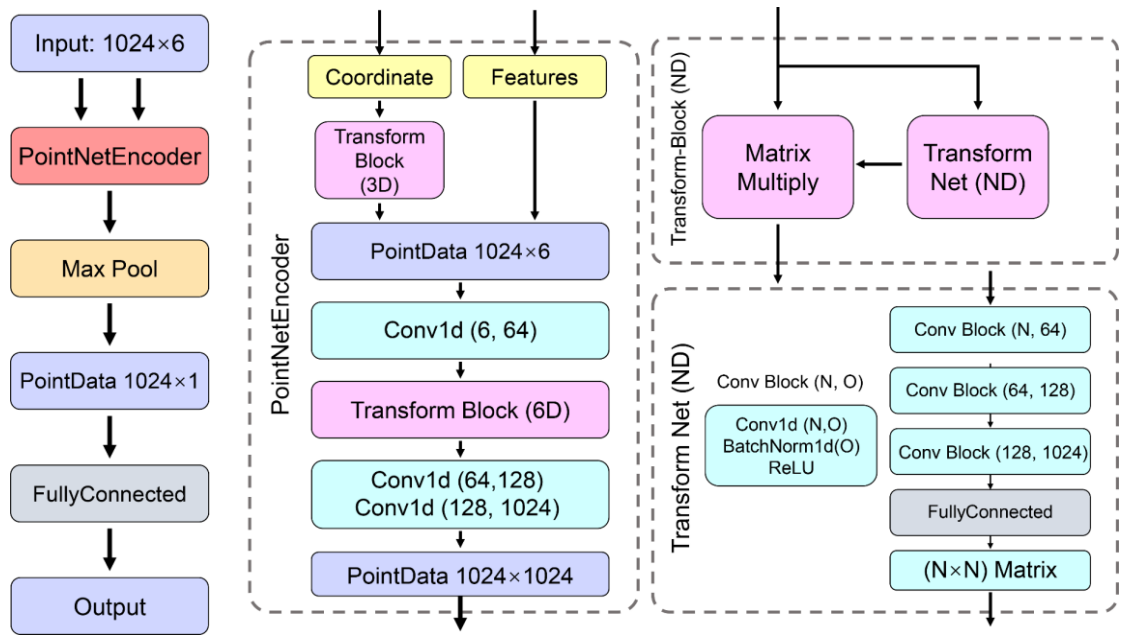

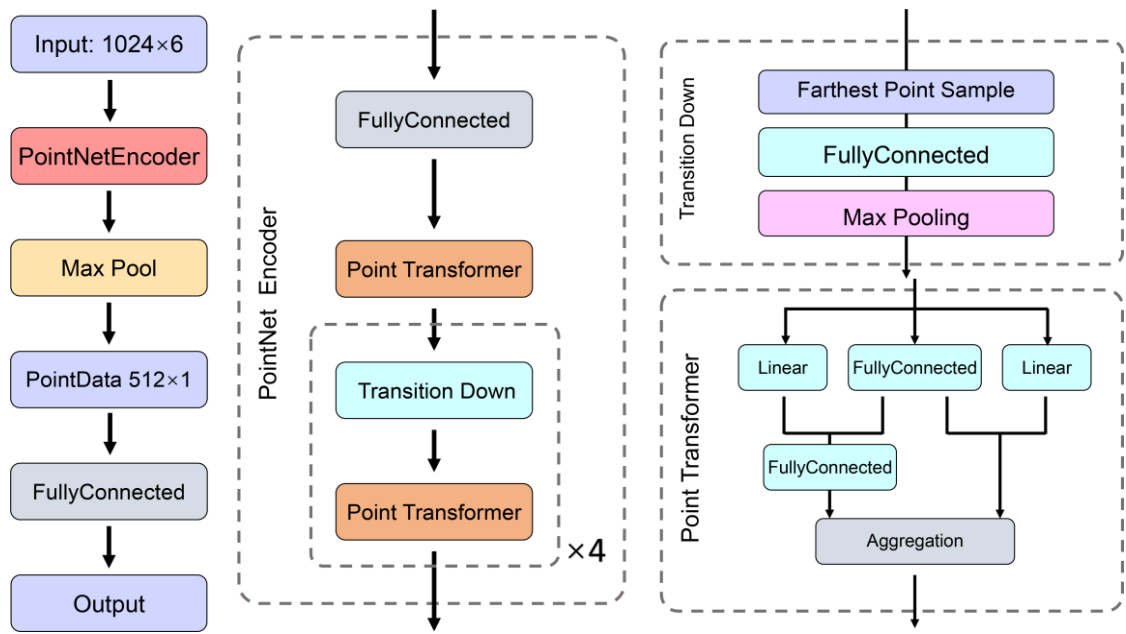

Figure S3. PointTransformer architecture.

**Figure S4.** A simple example to explain the meaning of Max Pool Layer. The atomic input order changes do not effect the output information after Max Pool.

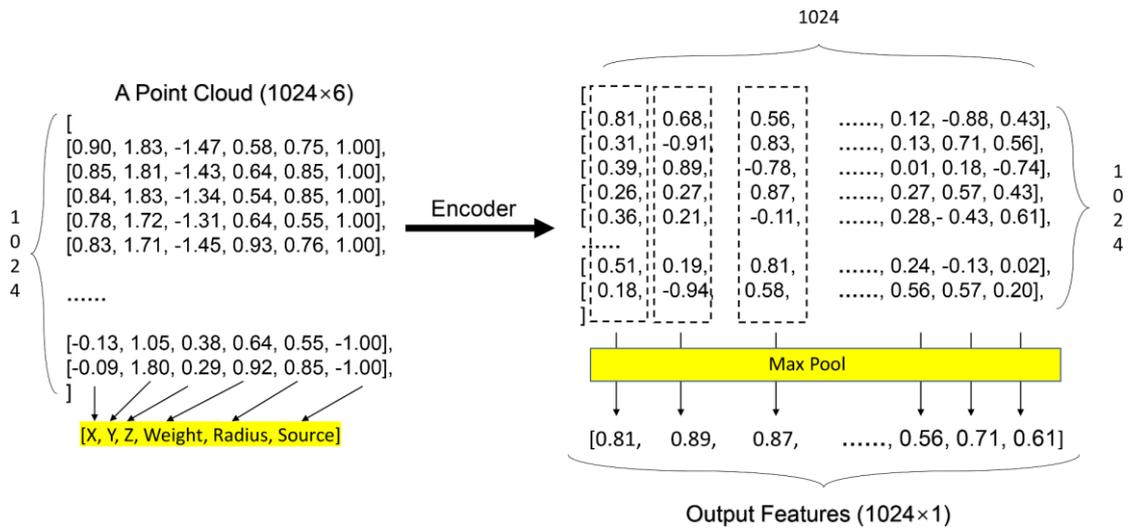

**Figure S5.** Visualization of the output results of the six protein-ligand complexs. The blue denotes the ligand, the yellow denotes the critical amino acid residue, and the red sphere denotes the atom that has the greatest effect on the prediction outcome. (a) 3L7B - PointNet (b) 3IH7 - PointNet (c) 2YKI - PointNet (d) 5C2H - PointNet (e) 4KZQ - PointTransformer (f) 1OYT - PointTransformer (g) 2YKI - PointTransformer (h) 5C2H - PointTransformer

# PointNet

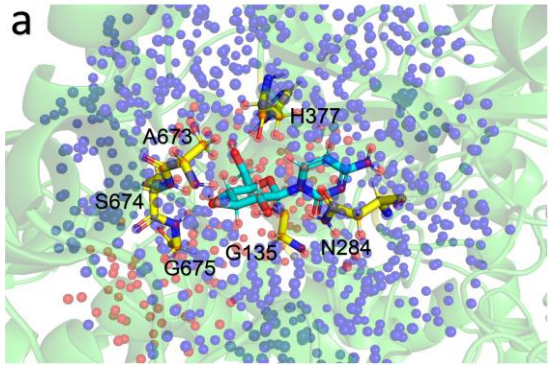

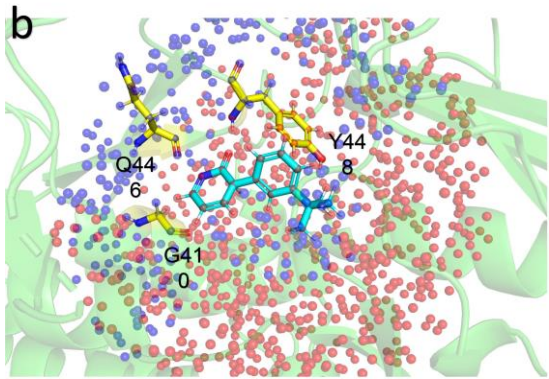

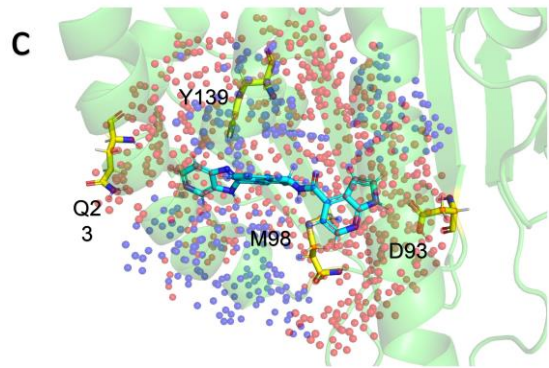

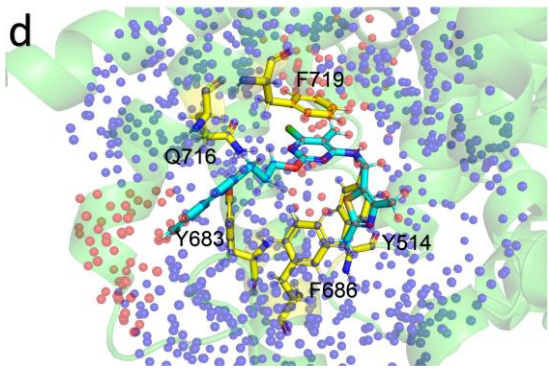

# PointTransformer

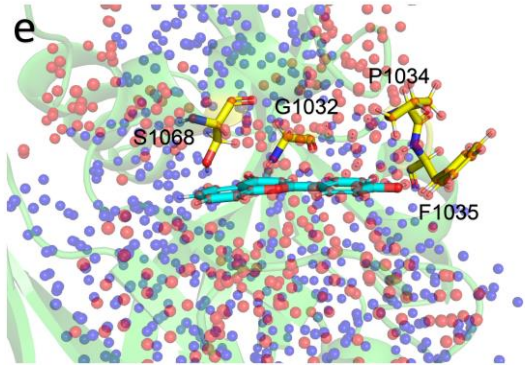

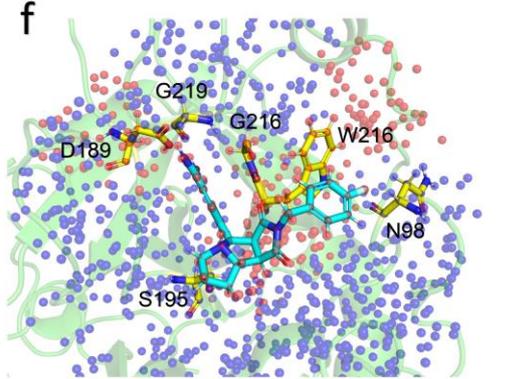

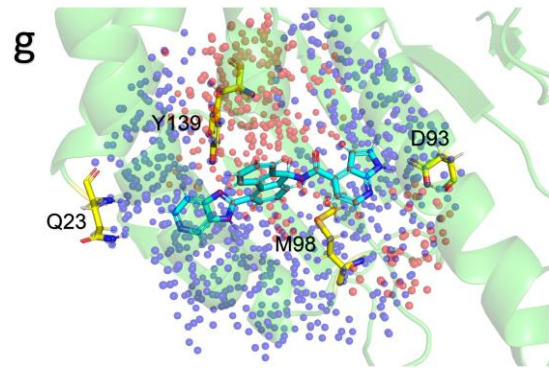

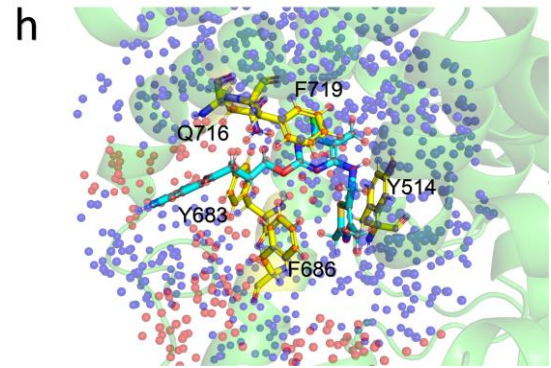

**Figure S6.** Visualization of 4KZQ by PointTransformer. The blue denotes the ligand, the yellow denotes the critical amino acid residue, and the red sphere denotes the atom that has the greatest effect on the prediction outcome.

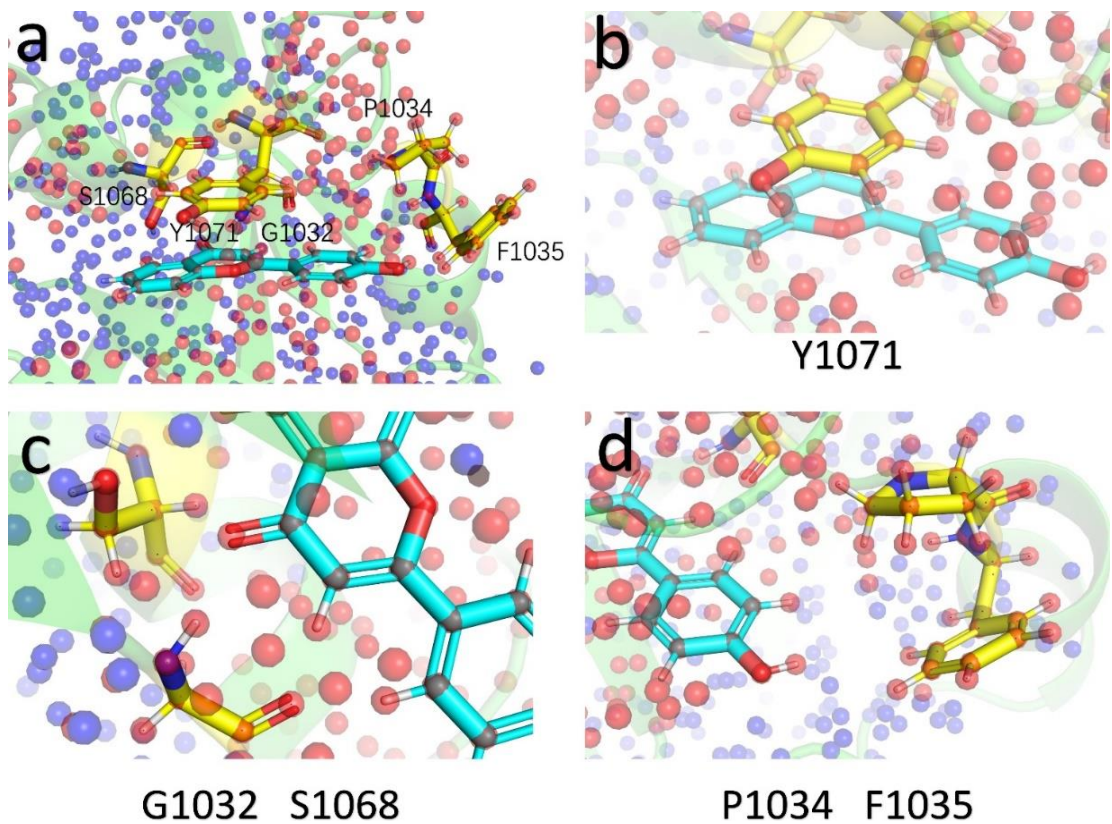

**Figure S7.** Visualization of 3L7B by PointNet. The blue denotes the ligand, the yellow denotes the critical amino acid residue, and the red sphere denotes the atom that has the greatest effect on the prediction outcome.

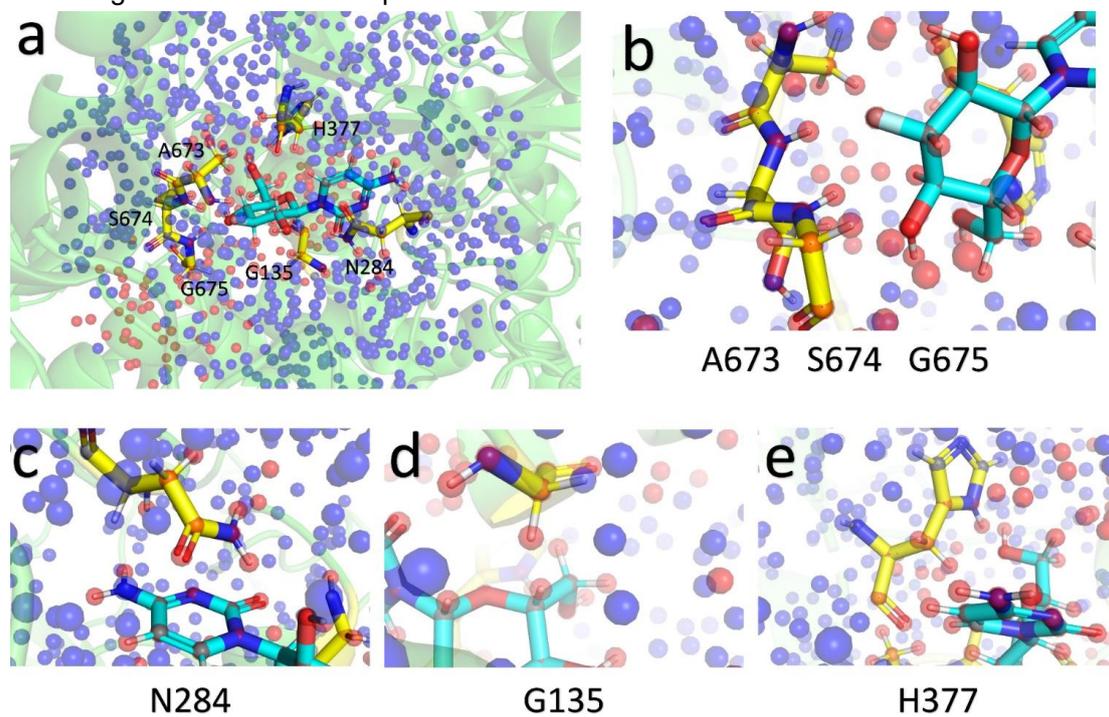